\documentclass[lettersize,journal]{IEEEtran}
\usepackage{amsmath,amsfonts}
\usepackage{algorithmic}
\usepackage{algorithm}
\usepackage{array}
\usepackage[caption=false,font=normalsize,labelfont=sf,textfont=sf]{subfig}
\usepackage{textcomp}
\usepackage{stfloats}
\usepackage{url}
\usepackage{verbatim}
\usepackage{graphicx}
\usepackage{cite}
\usepackage{listings}
\usepackage{xcolor}
\begin{document}

\title{Secure QR Codes: Authenticity Verification \\via EdDSA Signatures and CBOR Certificates}

\author{Wojciech Jonderko and \and Wojciech Wodo,~\IEEEmembership{Member,~IEEE}\\
\thanks{Manuscript received March 30, 2026; revised XXX XX, 2026.}
\thanks{W.Jonderko and W.Wodo are from Wroclaw University of Science and Technology, Wybrzeze Wyspianskiego 27, 50-370 Wroclaw, Poland, e-mail: wojciech.jonderko@pwr.edu.pl; wojciech.wodo@pwr.edu.pl}

}



\maketitle

\begin{abstract}
QR codes are a ubiquitous part of daily life, widely trusted by millions. However, their lack of inherent security features has given rise to critical attack vectors, such as spoofing (quishing) on public infrastructure like self-service parking machines. To address this, we present a comprehensive evolution of secure QR code architectures. First, we evaluate a fully offline proof-of-concept leveraging EdDSA signatures (instantiated on the Ed25519 curve), CBOR-encoded certificates, and ZLIB compression, demonstrating that robust cryptographic integrity can be achieved within the QR code's strict static capacity. However, recognizing the scalability limitations of fully offline models---specifically the inability to perform immediate key revocation in massive smart-city IoT deployments---we subsequently propose a scalable Hybrid Web PKI architecture. This forward-looking model utilizes standardized JWKS endpoints, a Central Trust Registry, and URL fragments to ensure seamless backward compatibility with standard native cameras while providing dynamic, real-time validation for compliant applications. This dual-mode approach offers a practical, deployable path toward eliminating QR spoofing 
\end{abstract}


\section{Introduction} \label{intro}

The widespread adoption of QR codes (Quick Response codes) has transcended their initial industrial purpose, embedding them deeply into the fabric of modern society. Today, they are ubiquitous in public spaces and digital environments alike. From restaurant menus and payment terminals to digital business cards, email signatures, and authentication prompts on computer monitors, QR codes serve as critical bridges between the physical and digital worlds. Their integration into banking apps and government services (e.g. COVID-19 certificates) has fostered a habit of reflexive scanning among users, who often treat them as benign shortcuts to information.

However, this convenience creates a significant security vulnerability known as "blind trust." Unlike a URL explicitly written in text, which a user might inspect for typos or suspicious domains, a QR code is visually abstract. A legitimate payment code and a malicious phishing link look functionally identical to the human eye. Users possess no natural ability to decipher the matrix barcode or verify its content before scanning. This inability to visually verify the origin or integrity of the data exposes users to "Quishing" (QR phishing) \cite{kicb2024quishing} and physical tampering attacks—such as criminals overlaying legitimate codes on parking meters with malicious stickers to divert payments \cite{fbi2022warning}.

The roots of this vulnerability are historical. Created in 1994 by Denso Wave, the QR standard was designed primarily for high-speed tracking of automotive parts in logistics. The priority was capacity and error tolerance, not security. Consequently, the standard lacks inherent cryptographic features; it includes no native mechanism for digital signatures, certificates, or issuer authentication. The data encoded is treated as trusted by default, leaving the validation burden entirely on the scanning application, which often simply executes the payload (e.g. opening a browser).

Existing literature does not sufficiently address the challenge of verifying the QR code issuer. While many studies focus on data hiding (steganography), visual aesthetics, or using AI to detect malicious redirects after scanning, there is a lack of standardized methods for both offline and online authenticity verification. To address this, we propose a secure QR code solution family leveraging established cryptographic primitives—digital signatures and lightweight certificates. Our approach assumes minimal changes to the user experience (UX): the scanning process remains effortless, but the system provides cryptographic proof of the issuer's identity and data integrity. Crucially, one of our solutions is designed to work offline and respects the storage constraints of standard QR versions, ensuring practical applicability in real-world scenarios where internet access may be intermittent or unavailable.

The core problem lies in blind trust: QR codes are visually identical whether legitimate or malicious, and users cannot visually verify their content or origin. 

\subsection{Motivation \& Contribution} \label{moti}

In recent years, the cybersecurity landscape has witnessed a surge in QR code-based attacks, a phenomenon increasingly referred to as "Quishing" (QR Phishing). Unlike traditional phishing via email, this attack vector exploits the physical layer of security. A prominent and growing threat involves \textbf{physical tampering}, where attackers overlay legitimate QR codes on public infrastructure—such as parking meters, electric vehicle charging stations, and bike-sharing terminals—with malicious stickers.

This vector has reached alarming intensity. \textbf{The National Institute of Cybersecurity (KICB) in Poland explicitly warns against these practices}, highlighting how fraudsters use fake QR codes to redirect users to malicious websites or fraudulent payment gateways \cite{kicb2024quishing, fbi2022warning}. These attacks succeed because users inherently trust the physical device (e.g. the city parking meter) and blindly extend that trust to the QR code adhered to it.

Despite the severity of these real-world incidents, existing literature does not provide a comprehensive solution for authenticating the \textit{issuer} of a QR code in an offline or online environment. Most current research focuses on:

\begin{enumerate}
    \item \textit{Data hiding and steganography} (embedding hidden data) \cite{ti2020visual, zhao2019qr, pan2022novel, hu2022threshold, yu2019three, lu2017multiple},
    \item \textit{Attack detection} (analyzing the destination URL for phishing patterns) \cite{alzahrani2021ai, mukhammedali2025ai},
    \item \textit{User authentication} (using the QR as a token, not verifying the QR itself) \cite{erg2023qr, reskita2023ecdsa}.
\end{enumerate}

None of these approaches effectively solve the problem of verifying the \textbf{origin} of the QR code itself before the payload is executed. The "blind trust" problem persists because users cannot visually decrypt the matrix to verify if it was signed by a legitimate city authority or a scammer.

\textbf{Contribution:} To address the challenge of "blind trust", we propose a secure QR code architecture leveraging EdDSA digital signatures (instantiated on the Ed25519 curve \cite{rfc8032}) and CBOR certificates. Our contribution is a lightweight, online and offline-verifiable "Chain of Trust" that empowers users to verify the identity of the signer (e.g. "City Parking Authority") and the integrity of the data. Key drivers of this project include maintaining minimal impact on User Experience (UX)—scanning remains fast and effortless—while ensuring the solution fits within standard QR capacity constraints.

\subsection{Structure}
This paper is structured as follows: Section II reviews related work and identifies current gaps in QR security research. Section III explains QR code standards, capacity constraints, and why traditional cryptography (RSA) fails in this context. Section IV presents our proposed solution, detailing the system architecture, cryptographic primitives (Ed25519/CBOR), and the "Chain of Trust" implementation. Finally, Section V discusses the experimental results, limitations of the current approach, and outlines future work.

\section{Related Work}
Existing research on QR code security addresses various aspects of the technology, ranging from data confidentiality to phishing detection. We categorize the state-of-the-art into four main groups, highlighting their limitations in the context of offline, public authentication.

\subsection{Heavy Cryptography (RSA)}
Early attempts to embed digital signatures in QR codes frequently relied on RSA. While cryptographically robust, standard RSA-2048 signatures (256 bytes) combined with X.509 certificates (often exceeding 1 KB due to ASN.1/DER encoding) create data payloads that surpass the practical capacity of standard QR codes. Implementing RSA forces the use of high-density QR versions (e.g. version 25 or higher), which require high-resolution printing and are difficult to scan with consumer smartphone cameras, particularly in poor lighting conditions or when the code is physically weathered \cite{evangeline2023rsa, erg2023qr}.

\subsection{Visual Secret Sharing and Steganography}
A significant body of work focuses on visual secret sharing schemes (VSS) and steganography. These methods often split the QR code into multiple shares or visually encrypt the matrix to hide sensitive data. While effective for privacy preservation and controlling access to the payload, these techniques do not solve the problem of public issuer authentication. They typically require the verifier to possess a pre-shared secret or a specific decryption key, making them unsuitable for public scenarios (like parking meters) where any random user must be able to verify the code's legitimacy without prior setup \cite{ti2020visual, zhao2019qr, pan2022novel, hu2022threshold, yu2019three, lu2017multiple}.
  
\subsection{Machine Learning and AI-based Detection}
Another prevalent approach utilizes Machine Learning (ML) and Artificial Intelligence to detect malicious QR codes dynamically. These systems analyze the decoded URL features or the visual characteristics of the QR code to identify potential phishing attempts. However, these solutions act as reactive filters rather than proactive proofs of authenticity. Furthermore, they predominantly rely on online connectivity to query centralized blacklists or reputation engines, offering limited to no protection in offline environments or against zero-day phishing domains \cite{alzahrani2021ai, mukhammedali2025ai}.

\subsection{ECDSA and Optimization Attempts}
Recognizing the size constraints of RSA, recent works have explored Elliptic Curve Digital Signature Algorithm (ECDSA) for QR authentication \cite{reskita2023ecdsa, malik2018ecdsa, gani2024ecdsa}. While ECDSA signatures are significantly smaller, the associated Public Key Infrastructure (PKI) based on X.509 certificates still introduces substantial overhead. To fit the data into a QR code, researchers often strip away the certificate entirely, embedding only the raw public key or signature. This compromise breaks the chain of trust, as it becomes difficult to verify whether the public key actually belongs to a trusted authority or an attacker, leading to scalable key management issues.

\subsection{Research Gap}
Currently, to our best knowledge, no solution reliably verifies the QR code's origin and issuer using a lightweight, offline-capable PKI that fits comfortably within low-density, easily scannable QR versions (e.g. version 15). Heavy signatures (RSA) exceed capacity, visual cryptography lacks public verifiability, and AI methods fail offline. Our work bridges this gap by combining EdDSA over the Ed25519 curve \cite{rfc8032} (for minimal signature size) with CBOR serialization, enabling a robust chain of trust without the bloat of traditional architectures. 

\section{QR Codes and Capacity Constraints} \label{qr_codes}

\subsection{Overview of the QR Encoding Standard} \label{qr_overview}
QR codes (Quick Response codes) are two-dimensional barcodes standardized in ISO/IEC~18004~\cite{iso18004}, designed to encode textual or binary data with fast decoding and high error tolerance. Unlike traditional 1D barcodes, they encode data in both vertical and horizontal dimensions, significantly increasing information density.

A QR symbol is not merely a random collection of pixels but follows a strict geometric structure essential for the scanning process. As illustrated in Figure~\ref{fig:qr_anatomy}, the symbol comprises specific functional regions \cite{infzak_qr}:

\begin{figure}[h!]
    \centering
    \includegraphics[width=1.0\linewidth]{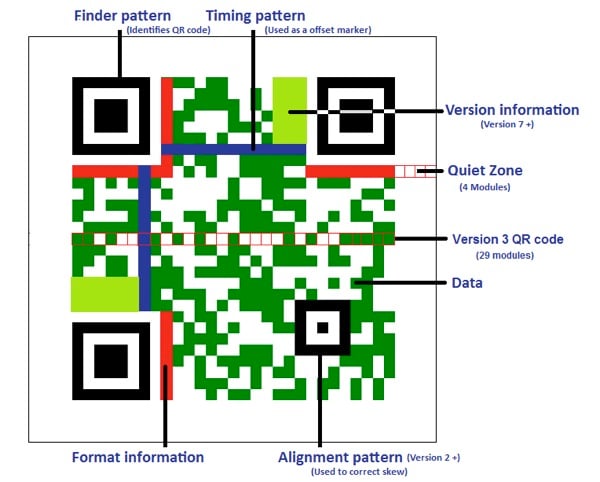}
    \caption{Anatomy of a QR Code symbol showing functional patterns, including version and Format Information regions (source: \cite{adafruit_qr}).}
    \label{fig:qr_anatomy}
\end{figure}

\begin{itemize}
    \item \textbf{Finder Patterns:} Three large concentric squares located at the corners (top-left, top-right, bottom-left) that allow the scanner to detect the code's position and orientation (360-degree readability).
    \item \textbf{Timing Patterns:} Alternating black and white modules connecting the finder patterns, serving as coordinate rulers to determine the module size.
    \item \textbf{Alignment Patterns:} Smaller isolated squares used to correct perspective distortion (skew) when the code is scanned at an angle. These appear in version 2 and larger.
    \item \textbf{Format Information:} Strips located adjacent to the finder patterns containing critical decoding parameters: the Error Correction Level (L, M, Q, H) and the Mask Pattern index.
    \item \textbf{Version Information:} Additional blocks located near the top-right and bottom-left finder patterns. These are present only in \textbf{version 7} and higher to ensure the scanner correctly identifies large, dense matrix sizes.
    \item \textbf{Quiet Zone:} A mandatory white margin of at least \textbf{4 modules} width surrounding the code to isolate it from visual noise.
    \item \textbf{Data Region:} The remaining area containing the actual payload intermixed with error correction codewords.
\end{itemize}

\subsubsection{Encoding Modes} \label{encoding}
The QR standard defines multiple encoding modes, selected automatically depending on the payload type to maximize density:

\begin{itemize}
    \item \textit{Numeric mode} – digits only (0-9), highest density (10 bits per 3 digits),
    \item \textit{Alphanumeric mode} – 45-character set (digits, uppercase letters, symbols), efficient for URLs,
    \item \textit{Byte/Binary mode} – raw 8-bit byte encoding (ISO/IEC 8859-1 or UTF-8),
    \item \textit{Kanji mode} – legacy Shift-JIS encoding.
\end{itemize}

For cryptographic payloads such as signatures, public keys, or certificates, \textbf{Binary Mode} is mandatory. Proper utilization of this mode is critical to avoiding unnecessary data expansion, as discussed in Section~\ref{binary_mode}.

\subsubsection{QR versions}  \label{qr-version}
QR codes are defined in 40 versions, each increasing the number of modules per side.
The symbol size grows linearly:
\begin{itemize}
    \item version~1: $21 \times 21$ modules,
    \item version~3: $29 \times 29$ modules (example shown in Figure~\ref{fig:qr_anatomy}),
    \item version~40: $177 \times 177$ modules.
\end{itemize}
Each version adds four modules per dimension. While higher versions offer greater capacity, they result in denser matrices that require higher resolution for printing and scanning. As noted in \cite{infzak_qr}, codes above version 7 require dedicated "Version Information" blocks, and versions exceeding 20 become difficult to scan in dynamic environments (e.g. handheld scanning of a parking meter).

\subsubsection{Error Correction Levels}  \label{error_correction}
QR codes use Reed-Solomon error correction with four predefined levels, which are stored within the "Format Information" region shown in Figure~\ref{fig:qr_anatomy}. Higher levels allow the code to be scanned even if part of it is damaged or obscured.

\begin{table}[h]
\caption{QR error correction levels \cite{iso18004}.}
\label{tab:qr_correction}
\centering
\begin{tabular}{|c|c|l|}
\hline
Level & ECC (\%) & Typical Use \\
\hline
L & $\sim$7\%  & Maximum capacity \\
M & $\sim$15\% & Balanced (default) \\
Q & $\sim$25\% & High reliability \\
H & $\sim$30\% & Robust physical labels \\
\hline
\end{tabular}
\end{table}

Higher error correction reduces available payload capacity, but is crucial for physical durability in public spaces, where codes are exposed to weather and vandalism \cite{infzak_qr}.

\subsubsection{Internal Data Structure} \label{internal_data}
Encoded data is arranged as a bitstream composed of:
\begin{itemize}
    \item mode indicator,
    \item character count indicator,
    \item encoded payload bits,
    \item terminator and padding bits,
    \item pad bytes (0xEC, 0x11 alternating),
    \item Reed-Solomon error correction blocks.
\end{itemize}

Before the final matrix is rendered, a \textbf{data masking} process is applied to the bitstream. The QR standard defines eight mathematical mask patterns that selectively invert modules (using an XOR operation) within the Data Region. The primary goal of masking is to break up large homogeneous areas of black or white pixels and to prevent the accidental formation of shapes resembling the Finder Patterns. The encoder evaluates all eight masks using a predefined penalty scoring system and selects the one that minimizes optical confusion. The index of the chosen mask is then recorded in the Format Information region (as mentioned in Section \ref{qr_overview}), enabling the optical reader to reverse the operation during decoding \cite{iso18004, infzak_qr}.

\begin{table}[h!]
\caption{Structure of the QR Code Data Bitstream (Byte Mode).}
\label{tab:bitstream_structure}
\centering
\begin{tabular}{|l|c|>{\raggedright\arraybackslash}p{4.2cm}|}
\hline
\textbf{Field} & \textbf{Size (bits)} & \textbf{Description} \\
\hline
Mode Indicator & 4 & Value \texttt{0100} selects Byte Mode \\
\hline
Character Count & 8--16 & Defines payload length (depends on version) \\
\hline
Data Payload & $8 \times N$ & The actual binary data (ISO-8859-1 or UTF-8) \\
\hline
Terminator & 4 & \texttt{0000} marks end of data string \\
\hline
Bit Padding & 0--7 & Aligns stream to 8-bit byte boundary \\
\hline
Byte Padding & Variable & Alternating bytes \texttt{0xEC}, \texttt{0x11} to fill capacity \\
\hline
\end{tabular}
\end{table}

To maximize efficiency and security, our solution replaces standard textual payloads with a structured binary format. Instead of simple ASCII strings, the QR code encapsulates a nested cryptographic container, effectively acting as a digital envelope (refer to the ``Matryoshka'' model in Section \ref{solution_concept}). This architecture mirrors the approach used in the \textbf{EU Digital Covid Certificate (UCC)}, which also relies on CBOR and digital signatures to ensure offline authenticity \cite{infzak_ucc, infzak_app}.

\subsection{Maximum Capacity of QR Codes} \label{qr_capacity}
The effective capacity of a QR code depends on its version, error correction level, and encoding mode.
Table~\ref{tab:qr_capacity} presents representative capacities for Byte Mode.

\begin{table}[h]
\caption{Maximum QR payload capacity in Byte Mode.}
\label{tab:qr_capacity}
\centering
\begin{tabular}{|c|c|c|c|}
\hline
Version & Bytes (M) & Bytes (Q) & Bytes (H) \\
\hline
10 & 213 & 151 & 119 \\
20 & 666 & 482 & 382 \\
26 & 1322 & 980 & 842 \\
\hline
\end{tabular}
\end{table}

Versions above 20 become physically large and difficult to scan in real-world conditions. Consequently, practical deployments typically target versions $\leq$ 15-27 to ensure compatibility with average smartphone cameras.

\subsection{Why RSA and ECDSA Are Not Suitable for QR Codes} \label{qr_rsa}

The limited capacity of QR codes imposes strict constraints on cryptographic material size. A secure QR code must fit within a scannable version (e.g. version 20) while maintaining a high error correction level (Q or H) to withstand physical damage.

\subsubsection{RSA-2048} \label{rsa}
A typical RSA-based package consists of:
\begin{itemize}
    \item RSA-2048 signature: 256~bytes,
    \item RSA public key: 256~bytes,
    \item X.509 certificate: 1000-1500~bytes,
    \item payload metadata: 50-200~bytes.
\end{itemize}

This results in a total size of approximately 1600-1800~bytes. Storing this data requires high-density QR versions (version 25+ for High ECC), which are prone to scanning failures in low light or at oblique angles \cite{evangeline2023rsa}. This makes RSA impractical for robust, public-facing QR codes.

\subsubsection{ECDSA (P-256)} \label{ecdsa}
ECDSA reduces key and signature size but still faces the certificate overhead problem:
\begin{itemize}
    \item signature: 64~bytes,
    \item public key: 64~bytes,
    \item X.509 certificate: 600-900~bytes.
\end{itemize}

The total size of 750-1000~bytes fits only in medium or large QR versions. While better than RSA, it leaves little margin for high error correction or additional payload data, limiting its flexibility in offline scenarios where the entire trust chain must be embedded.

\subsection{Why EdDSA (Ed25519) Is Well-Suited for QR Codes} \label{ed25519}
The EdDSA signature scheme instantiated on the Ed25519 curve \cite{rfc8032} provides compact, fixed-size cryptographic primitives that address the capacity bottleneck:
\begin{itemize}
    \item signature: 64~bytes,
    \item public key: 32~bytes,
    \item minimal CBOR certificate: 80-120~bytes.
\end{itemize}

With a typical payload of 40-120~bytes, the total QR package remains around 300~bytes. This comfortably fits in small QR versions (e.g. version 7-10) even with High (H) error correction. This property is decisive for creating secure, printable codes that are resilient to physical wear.

\subsection{Binary Mode as a Key Optimization} \label{binary_mode}
Many QR libraries default to encoding data as UTF-8 text, forcing binary data to be converted via Base64, hexadecimal, or JSON strings.
As detailed in \cite{infzak_base64}, Base64 encoding introduces a significant overhead of approximately 33\%, as it represents every 3 bytes of binary data with 4 ASCII characters. This increases the total size by 30-300\% depending on the serialization method.

Binary Mode allows raw byte storage without overhead. For example, a 64-byte Ed25519 signature remains exactly 64 bytes, whereas Base64 encoding expands it to 88 bytes.

The proposed system exploits Binary Mode by encoding a compressed CBOR payload. Specifically, the data processing pipeline follows three sequential steps: serialization via CBOR, compression using ZLIB, and final raw binary encoding. This specific data flow is strictly aligned with the technical specifications of the EU Digital COVID Certificate \cite{ehealth2021json, infzak_ucc}, which successfully deployed a similar architecture on a massive, international scale.

\section{Solution Concept} \label{solution_concept}

\subsection{Proof-of-Concept} \label{poc}
We developed a complete toolchain comprising key generation, certificate issuance, QR payload construction, and a verification reader. To provide operational flexibility, the system implements a \textbf{Dual-Mode PKI architecture}, allowing it to operate either via direct trust or a hierarchical certificate chain depending on the deployment scope. To foster transparency and encourage community adoption, the full reference implementation, including Python scripts for key management and verification, is available as open-source software on GitHub: [\url{https://github.com/bug-w4rri0r/qr-codes-security}].

\subsubsection{Data Structure ("Matryoshka" Model)}
\begin{figure}[h!]
    \centering
    \includegraphics[width=1.0\linewidth]{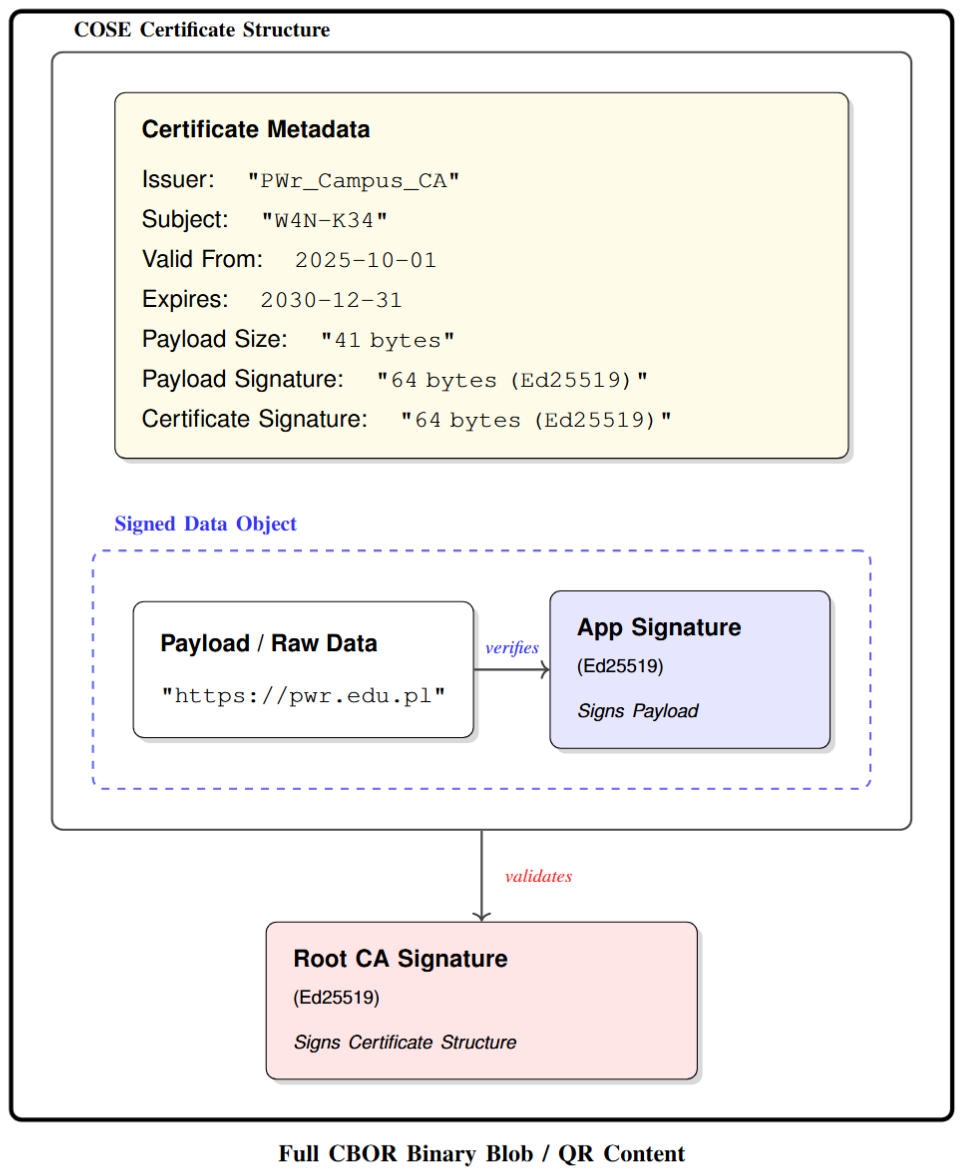}
    \caption{Logical structure of the "Matryoshka" model in Scenario 2. The entire payload and application identity are nested and cryptographically sealed by the Root CA signature.}
    \label{fig:matryoshka}
\end{figure}

To maximize efficiency, the system utilizes a nested data structure acting as a cryptographic envelope, which we refer to as the "Matryoshka" model. As illustrated in Figure~\ref{fig:matryoshka}, the trust is inherited from the outer layer (Root) to the inner layer (Payload).

The CBOR container adapts its structure based on the selected PKI mode:
\begin{itemize}
    \item \textbf{Mode A: Root-Only (Self-Signed / Direct Trust):} The hierarchical model is flattened. The payload is signed directly by the Root Private Key, effectively acting as a self-signed token without an intermediate certificate. The QR code contains only the user data and the signature (\texttt{\{payload, payload\_sig\}}).
    \item \textbf{Mode B: Root + App (Certificate Chain):} The full Matryoshka model is utilized. The payload is signed by the App, and the App's public key is encapsulated in a certificate signed by the Root. The QR code contains the full set (\texttt{\{payload, payload\_sig, cert, cert\_sig\}}).
\end{itemize}

The logical layers of the secure container are defined as follows, adhering to RFC standards for CBOR \cite{rfc8949} and COSE \cite{rfc8152}. Specifically, following the COSE specification \cite{rfc8152}, we utilize the EdDSA signature scheme instantiated on the Ed25519 curve \cite{rfc8032} for both signatures:
\begin{enumerate}
    \item \textbf{Payload:} The actual user data (e.g. a URL, Ticket ID, or Transaction Hash).
    \item \textbf{Payload Signature:} An EdDSA signature (Ed25519) \cite{rfc8032} of the payload created using the active App Private Key.
    \item \textbf{Lightweight Certificate:} A custom CBOR-encoded structure containing:
    \begin{itemize}
        \item \textit{Issuer ID} (referencing the Root CA),
        \item \textit{Subject ID} (identifying the device, e.g. "ParkingMeter-01"),
        \item \textit{Validity Period} (`nbf` - not before, `exp` - expiration),
        \item \textit{App Public Key} (used to verify the payload signature).
    \end{itemize}
    \item \textbf{Container Signature:} An EdDSA signature (Ed25519) \cite{rfc8032} covering the entire Certificate structure, signed by the offline Root CA.
\end{enumerate}

This binary structure ensures that the scanner can verify the Root's signature first. If valid, it trusts the enclosed App Public Key, which is then used to verify the actual payload.

\subsubsection{Packing Logic (Generator)}
The generation process transforms the raw data into a scan-ready QR code. While our architecture supports raw binary encoding for maximum density (as discussed in Section~\ref{binary_mode}), the Proof-of-Concept implementation utilizes an intermediate Base64 layer to ensure broad compatibility with standard text-based QR libraries during the testing phase.

The process, visualized in Figure~\ref{fig:creation_process_root} and Figure~\ref{fig:creation_process_chain}, involves the following steps:
\begin{enumerate}
    \item \textbf{Sign Payload:} The input data is signed with the App Private Key.
    \item \textbf{Construct CBOR:} A nested object is built: \texttt{\{payload, sig, cert, cert\_sig\}}.
    \item \textbf{Compress:} The CBOR bytes are compressed using ZLIB (Level 9) to reduce size \cite{rfc1950}.
    \item \textbf{Encode:} The compressed binary is Base64-encoded.
    \item \textbf{Generate QR:} The resulting string is rendered into a QR code (version dynamic, $\leq$ 26).
\end{enumerate}

\subsubsection{Verification Logic (Reader)}  
The verification follows a strict "fail-fast" philosophy:

\begin{enumerate}
    \item \textbf{Extraction:} Scan the QR code, decode Base64, and decompress ZLIB to retrieve the CBOR object.
    \item \textbf{Mode Auto-Detection:} The reader inspects the CBOR structure. If a certificate is present, it proceeds to Mode B validation; otherwise, it defaults to Mode A.
    \item \textbf{Step 1 - Trust Chain (Mode B only):} Verify the \texttt{cert\_sig} using the pre-installed, hardcoded Root Public Key. If validation fails, the code is rejected.
    \item \textbf{Step 2 - Metadata Check (Mode B only):} Validate that the current timestamp is within the certificate's \texttt{nbf} and \texttt{exp} window.
    \item \textbf{Step 3 - Payload Integrity:} Extract the Public Key (either the App Public Key from the validated certificate in Mode B, or the hardcoded Root Public Key in Mode A) and verify the \texttt{payload\_sig} against the payload.
    \item \textbf{Success:} If all checks pass, the authentic content is displayed along with the trust mode used.
\end{enumerate}

\subsection{Implementation Remarks} \label{implementation}
To ensure the system's security and practicality in real-world deployments, issuers must adhere to specific operational guidelines regarding key management and data optimization.

\subsubsection{Operational Key Management} \label{okm}
A distinct separation between the Root Authority and the Application level is critical.
\begin{itemize}
    \item \textbf{Root Private Key (Offline):} This key represents the ultimate trust anchor. It must be generated and stored in an offline environment (air-gapped) or within a Hardware Security Module (HSM). It is used \textit{only} to sign Application Certificates and must never be exposed to online field devices.
    \item \textbf{Application Keys (Online):} These keys are generated for specific devices (e.g. "Parking Meter \#1024"). If a physical device is compromised, its specific certificate can be revoked or allowed to expire without invalidating the entire system key set.
\end{itemize}

\subsubsection{Payload Optimization} \label{payload}
QR codes have a hard physical capacity limit. 
\begin{itemize}
    \item \textbf{URL Shortening:} Issuers should avoid encoding long, parameterized URLs directly (e.g. 200+ characters). Instead, use a domain-specific shortener (e.g. \texttt{https://city.gov/p/xyz}) should be used to keep the payload under 60-100 bytes.
    \item \textbf{Compression Strategy:} The built-in ZLIB compression is highly effective for text-based payloads (JSON, XML, URLs), typically reducing size by 30\%. However, for high-entropy data (random tokens), compression may be omitted to avoid overhead.
\end{itemize}

\subsection{Dual-Mode PKI Deployment Scenarios and PoC Evaluation}
The flexibility of the dual-mode architecture allows the solution to be tailored to specific operational environments. To validate our offline architecture, we executed both scenarios using our custom Python toolchain.

\subsubsection{Scenario 1: Root-Only (Direct Trust)}
For closed ecosystems or single-enterprise deployments, the hierarchical "Matryoshka" model is significantly flattened. The system utilizes a single key pair acting as a localized Trust Anchor. 

\begin{figure}[h!]
    \centering
    \includegraphics[width=1.0\linewidth]{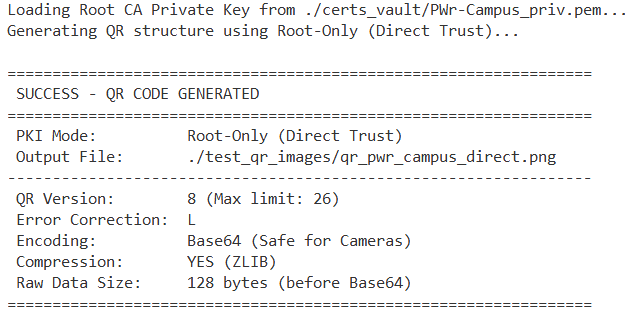}
    \caption{Terminal output demonstrating the Scenario 1 generation process: preparing the CBOR payload and signing it directly with the Root CA.}
    \label{fig:creation_process_root}
\end{figure}

\begin{figure}[h!]
    \centering
    \includegraphics[width=0.5\linewidth]{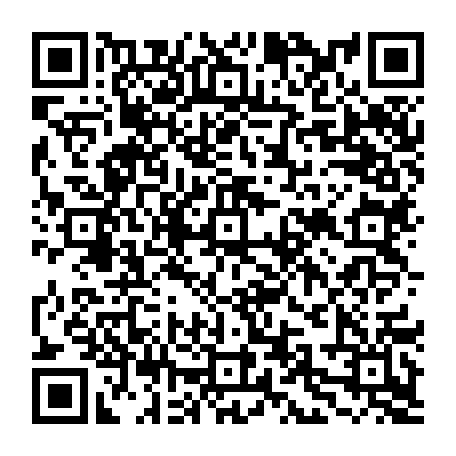}
    \caption{Scenario 1: Secure QR Code using Direct Trust. The low density (version 8) permits easy scanning.}
    \label{fig:poc_root_only}
\end{figure}

In our execution, the payload was signed directly by the Root CA. The generation process (Figure~\ref{fig:creation_process_root}) drastically reduces the CBOR payload size, producing a highly compact QR code, as shown in Figure~\ref{fig:poc_root_only} (version 8). To reproduce and verify this scenario, the following commands were executed:

\begin{lstlisting}[basicstyle=\ttfamily\scriptsize, breaklines=true, frame=single, backgroundcolor=\color{gray!10}, xleftmargin=0.5cm, xrightmargin=0.5cm]
# 1. Create QR using only the Root CA Private Key
python qr-create.py --payload "Welcome to the PWr campus (Direct Mode)!" --root-priv ./certs_vault/PWr-Campus_priv.pem --out ./test_qr_images/qr_pwr_campus_direct

# 2. Verify QR using the trusted directory (Auto-detects Mode A)
python qr-read.py --image ./test_qr_images/qr_pwr_campus_direct.png --trusted-dir ./trusted_roots/ --debug
\end{lstlisting}

\begin{figure}[h!]
    \centering
    \includegraphics[width=1.0\linewidth]{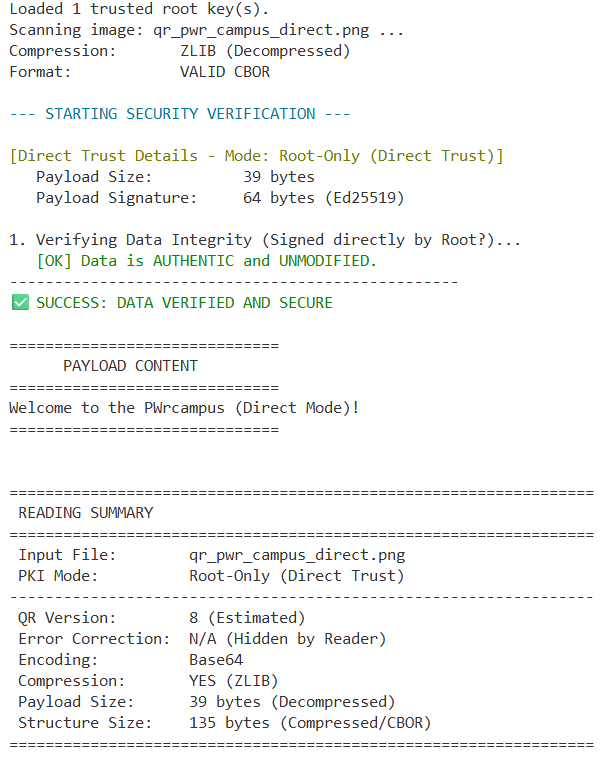}
    \caption{Terminal output demonstrating the Scenario 1 verification process: direct signature validation and payload extraction.}
    \label{fig:verification_process_root}
\end{figure}

Our terminal tests (Figure~\ref{fig:verification_process_root}) confirmed successful offline validation using only the pre-shared Root Public Key, without any intermediate certificates.

\subsubsection{Scenario 2: Delegated Trust (Certificate Chain)}
For large-scale, hierarchical environments, the full Certificate Chain is deployed. The Root CA issues expiration-bound certificates to specific departments. This allows the central authority to delegate signing capabilities without exposing the Root Private Key. 

In this scenario, the system employs the full Matryoshka model. The execution commands are as follows:

\begin{lstlisting}[basicstyle=\ttfamily\scriptsize, breaklines=true, frame=single, backgroundcolor=\color{gray!10}, xleftmargin=0.5cm, xrightmargin=0.5cm]
# 1. Create QR using the App Private Key and embedded CBOR Certificate
python qr-create.py --payload "Access to laboratory W4N-K34 granted." --app-priv ./certs_vault/app_priv.pem --cert-cbor ./certs_vault/cert_W4N_K34.cbor --out ./test_qr_images/qr_w4n_k34_chain

# 2. Verify QR (Auto-detects Mode B, verifies cert chain, then payload)
python qr-read.py --image ./test_qr_images/qr_w4n_k34_chain.png --trusted-dir ./trusted_roots/ --debug
\end{lstlisting}

\begin{figure}[h!]
    \centering
    \includegraphics[width=1.0\linewidth]{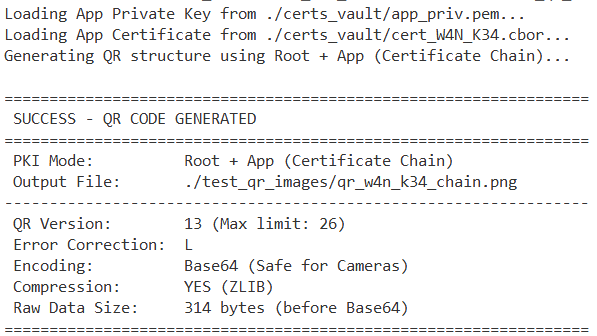}
    \caption{Terminal output demonstrating the Scenario 2 generation process: embedding the CBOR certificate and signing the payload with the App key.}
    \label{fig:creation_process_chain}
\end{figure}

\begin{figure}[h!]
    \centering
    \includegraphics[width=0.5\linewidth]{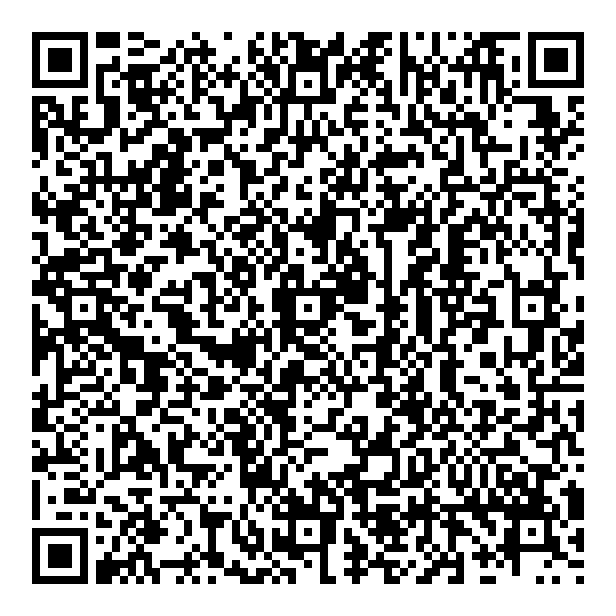}
    \caption{Scenario 2: Secure QR Code using a Certificate Chain. Increased density (version 13) accommodates the embedded CBOR certificate.}
    \label{fig:poc_root_app}
\end{figure}

\begin{figure}[h!]
    \centering
    \includegraphics[width=1.0\linewidth]{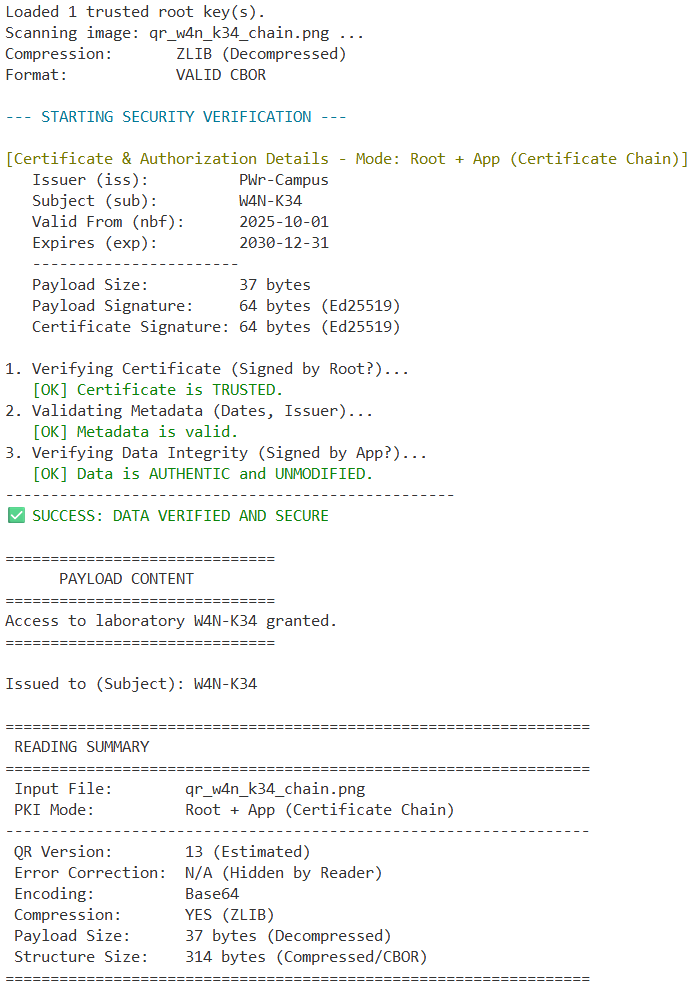}
    \caption{Terminal output demonstrating the Scenario 2 verification process: CBOR certificate parsing and trust chain validation.}
    \label{fig:verification_process_chain}
\end{figure}

Figure~\ref{fig:creation_process_chain} shows the generation phase, where the App's CBOR certificate is encapsulated within the payload. The resulting generated QR code is demonstrated in Figure~\ref{fig:poc_root_app}. While embedding the certificate increases the density (version 13), it enables offline trust delegation. The terminal output (Figure~\ref{fig:verification_process_chain}) successfully verified the App's certificate against the Root CA, followed immediately by the payload verification.


\subsection{Hybrid Architecture: Online Validation via Web PKI} \label{hybrid_architecture}

While our dual-mode offline evaluation proves that robust cryptographic integrity can be achieved without internet access, Scenario 2 highlights two inherent limitations for massive Smart City deployments. First, the physical footprint of the QR code grows significantly with the certificate chain. Second, executing an immediate, system-wide revocation of a compromised edge key remains challenging in a strictly offline regime.

To overcome these barriers, we propose an evolutionary step: the \textbf{Hybrid Trust Architecture}. In this model, the heavy certificate chain is offloaded from the QR payload to a trusted web server. The QR code encapsulates a standard URL---ensuring seamless backward compatibility with native smartphone cameras---appended with a cryptographic signature fragment.

Instead of embedding the public keys directly in the CBOR structure, the verification application parses the domain from the URL and securely fetches the authorized keys from standardized endpoints, such as \texttt{/.well-known/jwks.json} (JSON Web Key Set) or \texttt{/.well-known/security.txt}. These endpoints act as a Central Trust Registry, hosting dynamic key fingerprints. 

This hybrid approach drastically reduces the physical QR footprint (easily fitting within version 6 or 8) and enables real-time key revocation. It successfully bridges the gap, blending the cryptographic robustness of our offline proofs with the dynamic flexibility of modern Web PKI.

\subsubsection{Hierarchical Key Distribution (Web PKI)}
\begin{figure}[h!]
    \centering
    \includegraphics[width=1.0\linewidth]{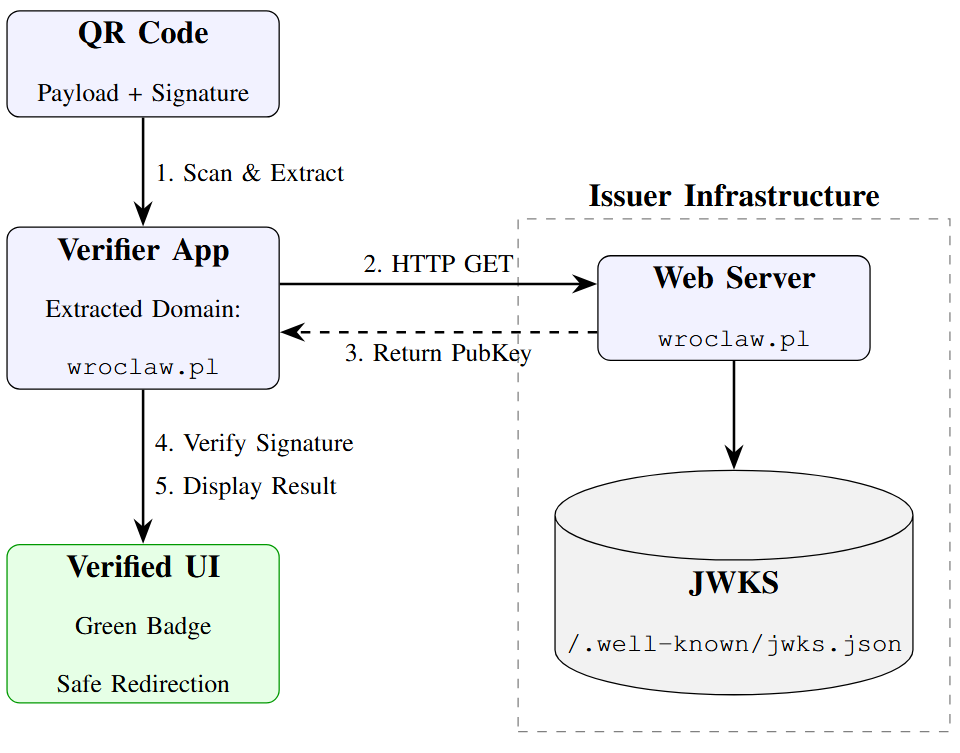}
    \caption{Hybrid Online Verification Flow. The application extracts the issuer's domain from the QR code and dynamically fetches the authorized public keys via the standardized JWKS endpoint.}
    \label{fig:hybrid_flow}
\end{figure}

To map the physical administration of public spaces to digital identities, the architecture employs a 3-tier Public Key Infrastructure (PKI), as illustrated in Figure~\ref{fig:hybrid_flow}:
\begin{enumerate}
    \item \textbf{Root Trust Anchor:} A central authority (e.g. a national digital agency or the developer of the official validation app). The Root Public Key is permanently embedded (hardcoded) within the mobile verification application.
    \item \textbf{Issuer (Sub-CA):} A municipal entity or service provider (e.g. the City Council). The issuer maintains an Intermediate Key pair, cryptographically signed by the Root.
    \item \textbf{Leaf Devices (Edge):} Individual hardware units (e.g. specific parking meters). These units hold only local private keys to sign the dynamic payload.
\end{enumerate}

\subsubsection{Standardized Key Distribution (JWKS)}
Instead of embedding the Issuer's public key in the QR code, the issuer publishes it on their official domain using the standardized \textbf{JSON Web Key Set (JWKS)} format \cite{rfc7517}. The keys are hosted at the universally recognized \texttt{.well-known} path \cite{rfc8615}, specifically: \texttt{\url{https://[issuer.domain]/.well-known/jwks.json}}. 

By utilizing JWKS, the system remains interoperable with existing web security infrastructures, avoiding proprietary distribution channels like \texttt{security.txt}, which is strictly reserved for vulnerability disclosure policies \cite{rfc9116}.

\subsubsection{Dynamic Verification Flow}
The verification process shifts the computational and data load to the smartphone's network layer. The flow is executed as follows:

\begin{enumerate}
    \item \textbf{Scan and Parse:} The smartphone scans the QR code, extracting the Payload (e.g. \texttt{https://wroclaw[.]pl/pay/123}), the cryptographic Signature, and a Key Identifier (\texttt{kid}).
    \item \textbf{Domain Extraction:} The application parses the host domain (\texttt{wroclaw[.]pl}) directly from the payload's URL. This eliminates the need to store the issuer's domain as separate metadata.
    \item \textbf{Fetch JWKS:} The app performs an HTTP GET request to \texttt{https://wroclaw[.] pl/.well-known/jwks.json}. To optimize performance, the app caches this file locally according to standard HTTP caching headers.
    \item \textbf{Trust Resolution:} Before trusting the fetched JWKS, the app verifies if the issuer's key was endorsed (signed) by the hardcoded Root Trust Anchor. This crucial step prevents DNS spoofing or attacker-controlled domains from injecting rogue keys.
    \item \textbf{Signature Validation:} Using the specific key matched by the \texttt{kid}, the app verifies the Ed25519 signature of the payload.
    \item \textbf{User Interface (UX):} Upon successful verification, the app displays a prominent "Verified Domain" badge (e.g. a green shield) along with the parsed URL, allowing the user to safely navigate to the payment gateway.
\end{enumerate}

This hybrid approach not only solves the capacity constraints of standard QR codes but also introduces real-time \textbf{Key Revocation}. If a specific parking meter is physically compromised, the administrator simply removes its corresponding public key from the \texttt{jwks.json} file on the city's server, instantly rendering any subsequently generated QR codes invalid.

\section{Conclusions and Future Work} \label{summary}

\subsection{Conclusions} \label{conclusion}
This paper presented a comprehensive architecture for ensuring the authenticity and integrity of QR codes in public spaces. By leveraging Ed25519 cryptography, CBOR serialization, and standardized web protocols, we demonstrated a robust approach to combat QR spoofing (quishing). The fully offline "Matryoshka" model provides secure, self-contained verification for air-gapped environments. The proof-of-concept validates that using modern, lightweight cryptographic primitives allows for a balance between strong security and the user convenience of quick scanning, successfully fitting the entire trust chain within the static capacity of a standard QR code.

\subsection{ Limitations} \label{limits}
While our proof-of-concept successfully validates the viability of offline verification using Ed25519 and CBOR, deploying this architecture in a real-world smart city environment (e.g. thousands of unattended parking meters) reveals distinct scaling challenges.

The primary limitation of the fully offline model is \textbf{immediate key revocation}. If a public infrastructure device is compromised and its private key is extracted, neutralizing the threat requires pushing a Certificate Revocation List (CRL) to every offline verifying device, which negates the offline advantage. Furthermore, encapsulating the entire certificate chain within the QR code rapidly consumes its data modules, restricting the use of lower, more reliable QR versions.

Additionally, a significant barrier to the widespread adoption of cryptographic QR codes is the absence of a dedicated, global Public Key Infrastructure (PKI) tailored for this medium. Currently, there is no centralized, universally trusted entity (Root CA) that issues certificates specifically for IoT devices on a smart-city scale. Consequently, the security model is vulnerable to human error, specifically \textit{typosquatting}. In an online scenario, if an attacker deploys a QR code containing a misspelled domain (e.g. \texttt{wr0claw.pl} instead of \texttt{wroclaw.pl}) and properly signs it with their own valid key hosted on that malicious domain, the cryptographic verification will succeed. The ultimate responsibility of verifying the semantic legitimacy of the URL still rests on the user's awareness. However, the Hybrid Web PKI architecture proposed in Section \ref{future_work} directly addresses and mitigates these exact limitations.

\subsection{Future Work: The Hybrid Web PKI Architecture} \label{future_work}
To overcome the limitations of the offline model, our future research will transition to a scalable, Hybrid Web PKI architecture. This approach shifts the trust anchor from an embedded CBOR certificate to a decentralized, online verification standard, drastically reducing the QR code payload to just the target URL and a cryptographic signature.

\begin{enumerate}
    \item \textbf{Backward-Compatible Payload Structure:} 
    We propose appending the Ed25519 signature as a URL fragment (e.g. \texttt{https://wroclaw.pl\#sig=[Base64]}). As illustrated in Figure~\ref{fig:compliant-flow}, this guarantees seamless backward compatibility. Standard cameras ignore the \texttt{\#sig} fragment and blindly redirect to the URL, whereas a Compliant Verifier App intercepts the fragment for validation.

    \item \textbf{Web PKI and Decentralized Key Fetching:}
    Instead of embedding certificates, the Compliant App extracts the domain from the URL and automatically queries standardized endpoints, specifically \texttt{/.well-known/jwks.json} (JSON Web Key Set) to fetch the issuer's current public keys, and \texttt{/.well-known/security.txt} to verify security policies. This guarantees \textit{immediate key revocation}---removing a compromised key from the JSON file instantly invalidates all spoofed codes city-wide.

    \begin{figure}[h!]
        \centering
        \includegraphics[width=1.0\linewidth]{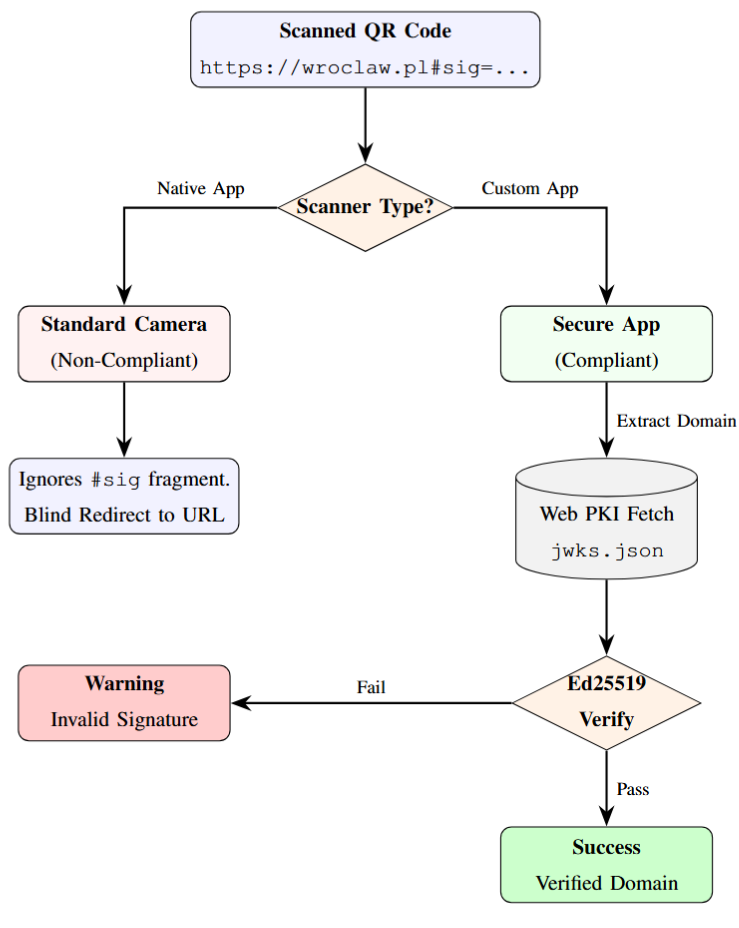}
        \caption{Proposed Backward Compatibility Flow. A standard camera ignores the signature fragment, while a compliant application utilizes decentralized Web PKI to verify the payload before redirection.}
        \label{fig:compliant-flow}
    \end{figure}

    \item \textbf{TLS Integration:} 
    Exploring methods to bridge the gap between traditional web certificates (X.509 TLS/SSL) and lightweight QR signatures. Future research will investigate deriving Ed25519 signing keys directly from existing TLS certificates (e.g. via Key Derivation Functions), allowing domains to leverage their existing TLS trust layer.

    \item \textbf{Central Trust Registry (Mitigating Domain Spoofing):}
    To mitigate the aforementioned typosquatting vulnerability (e.g. \texttt{fake-wroclaw.pl}), we propose integrating a Municipal or National Trust Registry. Before fetching the JWKS file, the verifier app cross-references the extracted domain against a securely updated whitelist of authorized government entities.
    
    \item \textbf{Mobile Integration and Caching:}
    To prevent verification latency on poor cellular networks, future mobile implementations will utilize aggressive, cryptographically secure local caching of JWKS files, bridging the gap between online security and offline speed.
\end{enumerate}


\bibliographystyle{IEEEtran}
\bibliography{references}


 



\vspace{-1cm} 

\begin{IEEEbiography}[{\includegraphics[width=1in,height=1.25in,clip,keepaspectratio]{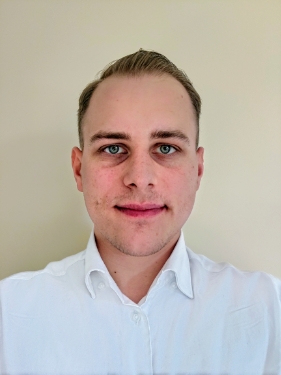}}]{Wojciech Jonderko}
received his B.S. in Telecommunications (2023) and M.S. in Cybersecurity (2024) from the Wroclaw University of Science and Technology, Poland. As a current Ph.D. candidate and Teaching Assistant at the same university, his research explores applied cryptography, PKI, and the security of IoT interfaces, focusing heavily on anti-spoofing mechanisms for QR codes. He actively contributes to the academic community as a co-founder of the IEEE Poland-Wroclaw Cybersecurity Group and a member of the university's cybersecurity team. Drawing on his broad industry experience, he successfully bridges theoretical research with practice. He applies his multidisciplinary expertise in designing scalable IT architectures, managing DevOps pipelines, and ensuring cloud infrastructure security to real-world enterprise solutions.
\end{IEEEbiography}

\begin{IEEEbiography}[{\includegraphics[width=1in,height=1.25in,clip,keepaspectratio]{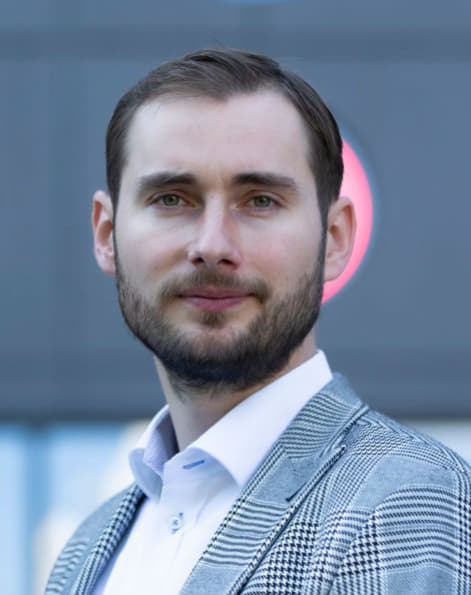}}]{Wojciech Wodo}
is an assistant professor at Wroclaw University of Science and Technology in Poland. He obtained his PhD degree from the Polish Academy of Sciences in Warsaw (2019) in the field of computer security. His areas of expertise are computer security and cybersecurity, especially in digital banking, electronic identity and biometrics fields. He is the author of two monographs on cybersecurity habits of digital banking users and the ecosystem of digital identity in Poland. Wojciech Wodo is also a member of the European Association for Biometrics and an experienced conference speaker - conducting talks on invitation at Volvo, UC Berkeley Haas School of Business, and the Trusted Economy Forum. He graduated from the Top 500 Innovators program at Haas School of Business UC Berkeley (2012) focused on science management, technology transfer and commercialization. Mr. Wodo created innovative teaching programs for students, being an author of two academic textbooks on entrepreneurship and innovation. He has served many times as a mentor and tutor for dozens of students over the last decade. Dr. Wodo loves teaching students, preparing his own lectures and classes on subjects such as biometrics or security of embedded systems. He supervises the scientific club "White Hats" at Wroclaw University of Science and Technology, which deals with cybersecurity, and cooperates on a regular basis with foreign research centers, including Lviv Polytechnic, Brno University of Technology, and Riga Technical University. Dr. Wodo is an Open Science enthusiast, engaged in EU international projects on openness, transparency and integrity in academia. He is also an Open Educational Resources practitioner and FAIR principles advocate.
\end{IEEEbiography}

\vspace{11pt}


\vfill

\end{document}